\documentstyle[11pt,newpasp,twoside,rotating,psfig,graphics]{article}
\index{central compact objects}
\index{central compact objects!associations with supernova remnants}
\index{magnetar}
\index{supernova remnants}

\def\asca{{\sl ASCA\/}}
\def\cxo{{\sl Chandra\/}}
\def\xmm{{\sl XMM\/}}

\def\edcomment#1{\iffalse\marginpar{\raggedright\sl#1\/}\else\relax\fi}
\reversemarginpar

\begin{document}
\title{Central Compact Objects in Supernova Remnants}
 \author{George G. Pavlov, Divas Sanwal \& Marcus A. Teter}
\affil{Pennsylvania State University, Department of Astronomy \&
Astrophysics, 525 Davey Lab, University Park, PA 16802}

\begin{abstract}
There are point-like sources in
central regions of several supernova remnants which have not been detected
outside the X-ray range. The X-ray spectra of these Central Compact Objects
(CCOs) have thermal components with 
blackbody temperatures of 
0.2--0.5 
keV and characteristic sizes of 
0.3--3 km.
Most likely, the CCOs
are neutron stars born in supernova explosions.  We overview their
observational properties, emphasizing the \cxo\ data,
and compare them with magnetars.
\end{abstract}
\vspace{-0.35in}
\section{Introduction}
X-ray observations have shown a new population of radio-quiet compact objects,
presumably isolated neutron stars (INSs), 
apparently different from both isolated rotation-powered pulsars 
and accretion-powered
X-ray pulsars in close binary systems.
About 10 of these objects
(AXPs and SGRs), which show pulsations with periods
in a 5--12 s range, large period derivatives, $\dot{P}\sim 10^{-12}$--$10^{-10}$,
and/or strong bursts are believed to be {\em magnetars} (Thompson \& Duncan 1996),
i.e., INSs with superstrong magnetic fields, $B\sim 10^{14}$--$10^{15}$ G. 
The magnetars
in quiescence have typical X-ray luminosities
$10^{34}$--$10^{35}$ erg s$^{-1}$ and
 show two-component spectra, a thermal component with a
blackbody (BB) temperature of 0.4--0.7 keV
and a non-thermal component 
described by a power-law (PL)
with a photon index around 3.
At least two magnetars have associated supernova remnants (SNRs).  
The second class
of radio-quiet INSs includes
colder (apparently older) nearby objects,  
showing soft thermal spectra with temperatures
in a 50--150 eV range and luminosities $\sim 10^{32}$--$10^{33}$ erg s$^{-1}$.
At least some of these objects, sometimes dubbed {\em Dim Isolated Neutron
Stars}, are likely
ordinary rotation-powered pulsars whose radio beams are not observable
from Earth.
Third is a class of point-like X-ray sources found near the centers of
SNRs that cannot be identified as active radio pulsars
or magnetars.  These {\em Central Compact Objects} (CCOs), presumably
INSs, have thermal spectra with BB temperatures  0.2--0.5 keV
and X-ray luminosities 
$L_{\rm x} \sim 10^{33}$--$10^{34}$ erg s$^{-1}$.  We define a CCO
as an X-ray point source which
(1) is found near the center of a SNR,
(2) shows no radio/$\gamma$-ray counterpart,
(3) shows no pulsar wind nebula (PWN),
(4) has a soft thermal-like spectrum.
Current observational status of CCOs is discussed below.

\vspace{-0.1in}
\section{Classification and General Observational Properties}

In the previous review by Pavlov et al.\ (2002a; P02 hereafter), 
five CCOs were discussed.
Since then three additional CCO candidates have been suggested
(Seward et al. 2003; Lazendic et al.\ 2003; Koptsevich et al.\ 2003).
Table~1
provides a list of the eight objects
(the new ones marked with `n'), their associated SNRs, the age
and distance to the SNR, the period, and the X-ray flux (in units of
$10^{-12}$ erg cm$^{-2}$ s$^{-1}$,
for a range of 0.4--8 keV). 
We believe that two of these objects, marked with `x' in Table 1,
do {\em not} actually belong to the CCO class because of their distinctive
properties, as discussed below.

\vspace{-0.15in}
\begin{table*}[ht]
\begin{center}
\begin{tabular}{lccccccccc}
\tableline
Object &  SNR & Age & $d$ & $P$ & $F_{\rm x,-12}$ \\
{}     & {}   & kyr & kpc & &   \\
\tableline
J232327.9+584843       & Cas A         & 0.32     & 3.3--3.7  & ...   & 0.8  \\
J085201.4$-$461753     & G266.1$-$1.2  & 1--3     & 1--2      & ...   & 1.4  \\
J161736.3$-$510225(x)  & RCW~103       & 1--3     & 3--7      & 6.4hr & 0.9--60 \\
J082157.5$-$430017     & Pup A         & 1--3     & 1.6--3.3  & ...   & 4.5  \\
J121000.8$-$522628     & G296.5+10.0   & 3--20    &  1.3--3.9 &424ms  & 2.3  \\
J185238.6+004020(n)    & Kes 79        & $\sim$9  & $\sim$10  & ...   & 0.2  \\
J171328.4$-$394955(n)  & G347.3$-$0.5  & $\sim$10 & $\sim$6   & ...   & 2.8  \\
J000256~~+62465\,\,\,(n,x) & G117.9+0.6[?] & $?$      & $\sim 3[?]$& ...  & 0.1 \\
\tableline\tableline
\end{tabular}
\caption{Compact Central Objects in supernova remnants.
}
\vspace{-0.2in}
\end{center}
\end{table*}


\noindent
{\em J1617--5102}:~~  This central object in RCW 103 was the first
radio-quiet X-ray point source found in a
young SNR (Tuohy \& Garmire 1980).  X-ray observations have shown that its
flux varies up to 2 orders of magnitude (Gotthelf, Petre \& Hwang 1997).
First \cxo\ 
observation showed a modulation of the light curve with about 6 hr period,
also found in the \asca\ data (Garmire et al.\ 2000).  In the second
\cxo\ observation the source flux had increased by a factor of 60.
The source has been monitored with \cxo\ ACIS, including a 50~ks observation
in Continuous Clocking (CC) mode which clearly showed a 6.4 hr period and 
multiple dips (Sanwal et al.\ 2002a).
The spectrum can be described by an absorbed BB,
with temperatures in a 0.4--0.6 keV range, anticorrelated with the flux.
The size of the emitting region varied from 0.2 to 1.6 km. 
The flux variability, the 6.4 hr period, and the dips in the X-ray light curve
suggest that this is {\em an accreting object in a binary system}.  
We observed the field of RCW 103 CCO in near-IR with 
VLT and {\em HST}/NICMOS and found
3 counterpart candidates within the 0\farcs7 Chandra error circle.  
The faintness of these objects (e.g., H=22--23)
 shows that the putative secondary companion
is a dwarf of a very late spectral type (later than M4).
Therefore, this source is
likely an unusual accreting binary, the second known binary in a SNR (after 
SS 443) and the first LMXB
in a SNR.  Since the other CCOs show no evidence of accretion,
we exclude this source from the CCO list.
\vskip 1.5mm
\noindent
{\em J0002+6245}:~~
This soft X-ray point source near the CTB~1 SNR, with a possible period of
242 ms, was discovered by Hailey \& Craig (1995).
The {\em ROSAT} observation showed a hint on a shell near CTB~1 which could be
a new SNR, G117.9+0.6, associated with this point source.
This source has been recently 
observed with {\sl XMM} for 30 ks (Koptsevich et al.\ 2003).
Very strong background flares 
left large parts of this observation unusable.  
The spectrum is best fit by a two-component model, a soft BB with
a temperature of 0.11 keV and a hard component, either a BB with temperature
of 0.5 keV or a PL with photon index of 2.6. The 242 ms period is excluded
by the {\sl XMM} observation, and no other periodicity is found.
No SNR around the source is seen in the {\sl XMM} images.  The spectral
parameters and a lack of a SNR suggest that the point source is a middle-aged
pulsar rather than a CCO. 

\begin{table*}[ht]
\begin{center}
\vspace{-0.2in}
\begin{tabular}{cccccccc}
\tableline
 Object & $kT$ & $R$ & $L_{\rm bol,33}$ & $\Gamma$ & $L_{\rm pl,33}$ & $n_{\rm H,22}$ & $F^{\rm bb}/F^{\rm pl}$  \\
   &  keV & km & & & &  & \\
\tableline
J2323+5848  
            & 0.43 & 0.6  & 1.6  & 4.2    & 13 &   1.8 & 1.1   \\
            & 0.43 & 0.7  & 1.9  & 2.5    & 0.2 & [1.2] & 4.5  \\
J0852--4617 & 0.40 & 0.3  & 0.3  & unconstr& ... &  0.4  & ...  \\
J0821--4300 & 0.40 & 1.0  & 3.3  & unconstr& ... & 0.3  & ...  \\
J1210--5226 & 0.22 & 2.0  & 1.2  & 3.6     & 1.2 &  0.13 & 3.0  \\
J1852+0040  & 0.50 & 1.0  & 8.0   & unconstr& ... &  1.5  & ...  \\
J1713--3949 & 0.38 & 2.4  & 15   & 3.9     & 72 &  0.8  & 0.9  \\
\tableline\tableline
\end{tabular}
\caption{Best-fit parameters for BB+PL fits to the {\sl Chandra}
spectra.
}
\end{center}
\vspace{-0.22in}
\end{table*}

The spectral parameters 
for the remaining six CCOs, obtained from fits to \cxo\ ACIS spectra,
are listed in Table~2 for 
the two-component BB+PL model, which became a standard description for
INS spectra.  The bolometric luminosities 
$L_{\rm bol}$ for the BB components, and the luminosities $L_{\rm pl}$ 
for the PL components (in the 0.4--8 keV band), 
are in units of $10^{33}$ erg s$^{-1}$. 
The hydrogen column densities are in units of $10^{22}$ cm$^{-2}$.
Significant contributions to the spectra
are given
by thermal-like emission, 
with BB temperatures 0.2--0.5 keV and BB radii
0.3--2.4 km, smaller than the expected NS radii, $R_{\rm NS}=10$--15 km. 
Fits with the light-element (H or He) atmosphere models (Pavlov et al.\ 1995)
 give lower temperatures, by a factor of $\sim 2$, and larger radii,
by a factor of 2--7, 
but still the radii remain $<R_{\rm NS}$ for at least two CCOs (see P02).
The PL component
is unconstrained in at least some of the fits, which means that either
the corresponding spectra are purely thermal or the PL component is
too faint to be detected in these observations. Fits with the one-component
PL model yield very steep slopes, $\Gamma \sim 5$ (see P02),
and they are statistically unacceptable in the cases when many source counts
were collected. Since the atmosphere spectra are harder than the BB spectra
in the X-ray band, the PL components are, as a rule, unconstrained in the
atmosphere+PL fits. 
Among the six CCOs, only one, J1210--5226, clearly shows spectral lines
(see below) while the others are satisfactorily described by featureless
continua. The J1210--5226 is also the only CCO for which a period 
has been detected. None of the six CCOs has shown any long-term variability.

\vspace{-0.1in}
\section{Individual Sources}

Here we describe the properties of the individual CCOs, with main emphasis on
the new results obtained in the last two years. More details about the previous
results and the references to earlier works can be found in P02.
\vskip 1.5mm
\noindent
{\em Cas A CCO:}~~
This prototype CCO has been observed many times with {\sl Chandra}
since its discovery in the first-light {\sl Chandra} observation
(Tananbaum 1999).
We have analyzed the archival {\sl Chandra} observations 
to search for periodicity, look for long-term variability, and determine
its spectral properties (Teter et al.\ 2003, in preparation).
We searched for a period of the point source using 
two 50 ks HRC observations 
(Dec 1999 and Oct 2000) and two 50 ks ACIS observations (Jan 2000 and
Feb 2002).  
No significant periods were found between 0.01 and 100 Hz.
To search for long-term variability,
we additionally used the calibration observations
done every
6 months (typical exposures 1--3 ks) and found no statistically significant
variations.
The lack of variability during 4 years of {\sl Chandra} observations
suggests that it is {\em not an accreting object} (hence 
{\em not a black hole}).

Spectral analysis was performed using the two 50 ks ACIS-S observations 
and the 70 ks HETG/ACIS observation of May 2001.
We find that the 
BB model gives a better fit than the PL model, with 
$kT =0.46\pm 0.01$ keV and 
$R=0.58\pm 0.03$ km.
The two-component models
(BB+PL and BB+BB) provide significant improvements over the BB fit,
with the soft-BB temperature ($0.43\pm0.02$ and $0.37\pm0.03$ keV)
 and radius ($0.6\pm 0.1$ and $0.8\pm 0.1$ km) similar to those obtained
from the single-component BB fit. 
The photon index of the PL component is strongly correlated with
the $n_{\rm H}$ value: $\Gamma =4.2\pm0.2$,  $n_{\rm H,22}=1.8\pm0.1$
if $n_{\rm H}$ is a free parameter, while $\Gamma=2.5\pm0.3$ for a more
realistic (fixed) $n_{\rm H,22}=1.2$.
The hard-BB component has a
temperature of $0.6\pm 0.1$ keV
 and 
a radius of $0.2\pm0.1$ km, corresponding to $n_{\rm H,22}=1.2\pm0.1$.
Substituting
H atmosphere models for the thermal emission gives somewhat better fits,
with lower temperatures and larger radii (e.g., $0.33\pm0.01$ 
keV and $2.2\pm 0.2$ km for a single-component low-field 
H atmosphere model), but the radii are still well below $R_{\rm NS}$.
The parameters of the hard component in two-component fits are not
very sensitive to the choice of soft component model and are less constrained
when the H atmosphere models are used.
Overall, we conclude that the mostly thermal radiation of the Cas A CCO
is emitted from a small, hot area ($R<2$ km, $kT>0.3$ keV), perhaps
a hot spot on the NS surface, and it is not associated
with accretion.
\vskip 1.5mm
\noindent
{\em CCO in ``Vela Junior''}:~~
The SNR G266.1--1.2 was discovered by Aschenbach (1998) in the south-east
corner of the Vela SNR in the RASS data. 
An imaging observation with {\sl Chandra} allowed
Pavlov et al. (2001) to detect a point source $4'$ from the SNR
center and 
measure
its position. 
The limiting optical magnitude B$>$22.5 gives high enough
X-ray-to-optical flux ratio to believe that this source is a NS,
possibly the remnant of the SN explosion.
The spectrum of the source was measured by Kargaltsev et al.\ (2002) from 
a 30 ks \cxo\ ACIS observation in the CC mode.
The PL model does not fit, while the BB model fits very well,
 giving 
$kT\simeq 0.4$
keV and 
$R\simeq 0.3$ km, assuming a distance of 1 kpc.
No significant pulsations are found from these data.
A 25 ks {\sl XMM} observation 
(Becker \& Aschenbach 2002) gave very similar
parameters for the thermal-like radiation and a poorly constrained
PL component ($\Gamma = 2.85\pm1.0$, $F^{\rm pl}/F^{\rm bb}\approx
0.15$ in the 0.5--10 keV band). 
Based on the observed  properties,
we conclude that this source is of the same nature as the Cas A CCO.
\vskip 1.5mm
\noindent
{\em Pup A CCO:}~~
J0821$-$4300, located about $6'$ from the center of Puppis A,
was discovered with {\sl Einstein} (Petre et al.\ 1982) and studied with
{\sl ROSAT}, {\sl ASCA}, {\sl Chandra} and {\sl XMM}.  
Its X-ray spectrum is very similar to those of the Cas A and Vela Junior
CCOs. The spectrum observed with {\sl Chandra} fits well
with one-component thermal (BB or light-element atmosphere) models
(P02),
while fitting
the {\sl XMM} spectrum requires an additional hard component
(PL with $\Gamma=2.0$--2.7 or hard BB with $kT=0.5$--1.1 keV ---
Becker \& Aschenbach 2002).  
This CCO has a larger size of the emitting region ($R\sim 1$ km 
and $\sim 10$ km for the BB and magnetic H atmosphere models, respectively).
Observations 
with {\sl Chandra} HRC have
shown no PWN
and no significant
periodicity in the 0.003--300 s range (P02).
\vskip 1.5mm
\noindent
{\em CCO in PKS 1209--51/52:}~~
J1210$-$5226
(a.k.a. 1E~1207.4--5209) 
was discovered 
by Helfand \& Becker (1984).
It is located about $6'$ off the
center of PKS~1209--51/52 (G296.5+10),
at a distance of about 2 kpc.  This source has been observed
with all the X-ray observatories since {\sl Einstein} (see P02
for references).
The low-resolution spectra obtained in the
pre-{\em Chandra} era
can be described by thermal continuum models; e.g, the BB fits give
$kT\simeq 0.25$ keV and $R\simeq 1.6$ km.
Fits with magnetic NS atmosphere models
show a lower temperature and a size compatible with
that of a NS (Zavlin et al.\ 1998).

Two {\sl Chandra} observation of this source resulted in the discovery of
a period of 424 ms
(Zavlin et al.\ 2000)
and a surprisingly small period derivative, corresponding to
a characteristic pulsar age of $\sim 500$ kyr (vs.\ 3--20 kyr for
the SNR) and magnetic field $\sim 3\times 10^{12}$ G (Pavlov
et al.\ 2002b).  Spectral
fits to the \cxo\ data show two broad absorption lines, near 0.7 and 1.4 keV
(Sanwal et al.\ 2002b), the first lines detected in an INS
spectrum.  Further observations of this
source with \xmm\ have shown that there might be additional absorption
lines near 2.1 and 2.8 keV (Bignami et al.\ 2003).  The origin of the lines
is not clear at present.

We found no long-term flux variations, neither between different observations
nor within long separate observations, which suggests that the X-ray
radiation is {\em not} associated with accretion.  On the other hand,
timing analysis of the \cxo\ and \xmm\ observations have allowed us to
discover apparent deviations from uniform spin-down, suggesting that the 
CCO could be in a wide binary system, with $P_{\rm orb}\sim 0.2$--6 yrs
(Zavlin et al.\ 2003).
We observed the field with VLT
and {\sl HST}/ACS and found a faint, red object (V$\simeq$26.4, K$_s\simeq$20.7)
in the $1''$ {\sl Chandra} error circle,
whose spectrum suggests an M4 or M5 dwarf (Moody et al.\ 2003, in preparation).
If
this is the CCO optical counterpart,
then it would be the second LMXB (but not a usual one!)
in a SNR, after the central source in RCW 103.

Thus, we see that J1210--5226, the coldest (the oldest?)
among the six CCOs, is the only one showing spectral lines
(but none of the others was observed for so long time),
the only one for which a period was detected (albeit with a puzzling
time dependence), and the only one that might be in a (non-accreting)
binary. Solving the riddles exhibited by this best-studied CCO
may give a clue to understanding the nature of the enigmatic CCO family,
unless  this ``outstanding CCO'' is a truly unique object.
\vskip 1.5mm
\noindent
{\em CCO in Kes 79:}~~
Kes 79 (G33.6+0.1) is a shell-like SNR (diameter $\sim 11'$)
at a distance of $10\pm2$ kpc, with an age of 6--12 kyr (Seward \&
Velusamy 1995). 
It has been observed in X-rays with {\sl Einstein},
{\sl ROSAT},
and {\sl ASCA}.
{\sl Chandra} ACIS observation of Kes 79 
(Seward et al.\ 2003) showed
a point-like source at its center.
Since neither PWN nor optical/radio counterparts have been detected,
this source is a viable CCO candidate. 
The upper limit on pulsed fraction is about 30\%,
for periods longer than 6.4 s.
The spectrum of this putative old CCO is thermal-like,
its BB temperature, 0.48 keV, being even higher than that of the youngest
Cas A CCO, is comparable to magnetar temperatures of similar ages. 
\vskip 1.5mm
\noindent
{\em CCO in G347.3--0.5:}~~
G347.3--0.5 is a radio-faint, shell-like SNR
with a nonthermal X-ray spectrum,
at a distance of about 6 kpc.
{\sl ROSAT} PSPC  observations of G347.3--0.5
(Pfeffermann \& Aschenbach 1996) showed a central point source,
for which neither optical/radio counterpart nor X-ray pulsations
have been detected (Slane et al.\ 1999).
Results of the {\sl Chandra},
{\sl XMM} and {\sl RXTE} observations of this source have been reported
by Lazendic et al.\ (2003).
The combined \cxo\ and {\em XMM} spectra of the CCO can
be reasonably fit with either a single BB ($kT \simeq 0.4$ keV, 
$R\simeq 2.5$ km)
or a steep PL ($\Gamma \simeq 4.2$).  
The best fit to the CCO spectrum is given by a 
BB+PL model, 
with $kT=0.38$ keV and 
$\Gamma = 3.9$, similar to
other CCOs where the two-component model fits are constrained.

\vspace{-0.1in}
\section{Evolution and H-R diagram for CCOs and Magnetars}
The above-described spectral observations of CCOs make it possible
to examine the age dependence of the spectral parameters.
Figure 1 shows
such dependences for the BB temperature and radius. Since the observed CCO 
spectra are similar to those of magnetars in quiescence, we added the data
on a few AXPs and SGRs (from Mereghetti et al.\ 2002).
\vspace{-0.7cm}
\begin{figure}[ht]
\centerline{
\hbox{
\vspace{-1.0cm}
\psfig{file=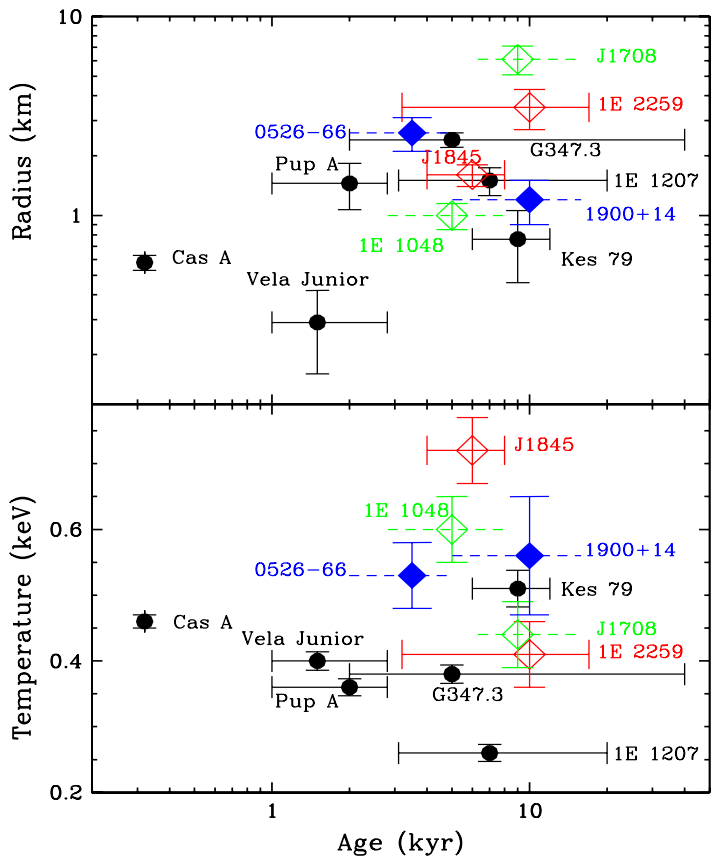,angle=0,width=3.9in,height=4.3in}
\hspace{-0.4in}
\parbox[b]{48mm}{
\caption{
Age dependences of BB radius and BB effective temperature
for CCOs (filled circles), AXPs (open diamonds)~and SGRs (filled diamonds).
For objects with~known SNR~associa\-tion,
esti\-mated SNR ages are used (solid hori\-zon\-tal error\-bars)
while for four magne\-tars (marked with~horizon\-tal da\-sh lin\-es) we use
spin\--down ag\-es.
}}
}}
\vspace{0.25in}
\end{figure}
We see from Figure 1 that CCOs, being
on average younger than magnetars, have somewhat lower temperatures
and smaller sizes, albeit with significant overlaps. 
Interestingly, the overlap in the temperature-age
diagram disappears if
the Kes 79 CCO is attributed to the magnetar group. In this case
the five CCOs lie along a ``cooling branch'', while the magnetars show
no temperature-age correlation. If we consider CCOs and magnetars as 
a single group, no significant temperature-age correlation is seen.
On the other hand, the effective radii of CCOs and magnetars 
do show positive correlation with age (a hot spot spreads over the
NS surface? a hole in a `screen' gets bigger?),
with magnetars being on average older 
than CCOs.
In the luminosity-temperature diagram
(an analog of Hertzsprung-Russell diagram --- Fig.\ 2)
the CCO and magnetar populations almost do not overlap (magnetars are
hotter and more luminous), lying on the same `sequence'.

\begin{figure}[ht]
\vspace{-0.6cm}
\centerline{\hbox{
\vspace{-2.5cm}
\hspace{0.5cm}
\psfig{file=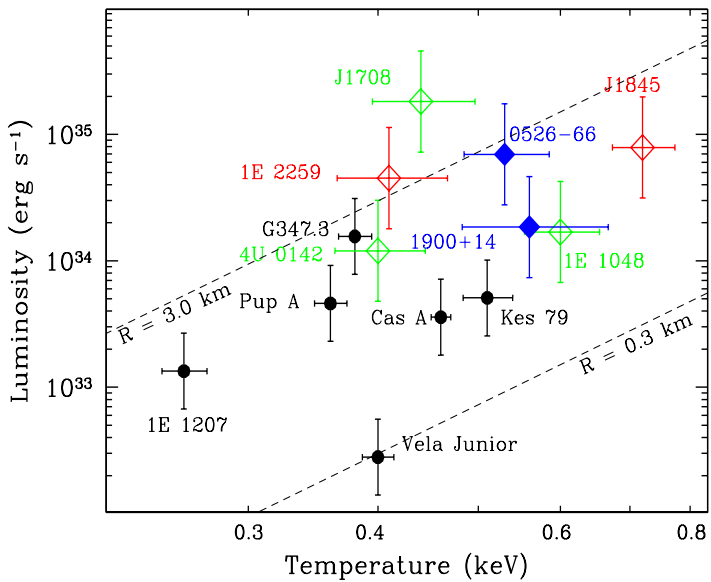,angle=270,width=3.5in,height=2.8in}
\hspace{-0.4in}
\parbox[b]{50mm}{
\vspace{-0.8in}
\caption{
The dependence of bolometric BB luminosity on BB temperature
(Hertzsprung-Russell diagram) for CCOs and magnetars.
The lines of
constant BB radius are plotted at 0.3 km and 3.0 km.
}}
}}
\vspace{0.32in}
\end{figure}

Thus, at least some CCOs appear to be relatives of AXPs and SGRs,
but their relationship is not fully understood. On average, CCOs are
younger, perhaps colder, and their emitting areas are smaller than those
of AXPs and SGRs. Does it mean that CCOs 
are actually young magnetars, not mature enough to develop
characteristic properties of AXPs/SGRs (e.g., they have not spun down
to the 5--12 s period range, or their crusts are still too durable
to crack and cause bursts)? This hypothesis could, at least, explain
the lack of pulsar activity in CCOs, but it seems to be at odds
with the properties of 1E~1207.4--5259,
the best-studied CCO (unless it is a different kind of object).

To conclude, we now have strong reasons to believe that CCOs are
not black holes, and their radiation is not powered by accretion.
Very likely, they are isolated ``neutron stars'' (perhaps composed
of more exotic particles, e.g., quarks),
but at least most of them are not ordinary rotation-powered pulsars.
Apparently, their thermal-like X-ray emission emerges
from a part of NS surface.
We can speculate that the internal heat of the NS is somehow
channeled into these small areas (e.g., by superstrong localized magnetic
fields) or most of the surface is covered by a thermo-isolating
``blanket'' or a ``screen'' opaque for soft X-rays. Alternatively,
the hot spots could be heated by some local sources (dissipation of
superstrong magnetic fields? nuclear reactions?). 
Future X-ray timing and spectral observations, together with 
deep NIR imaging, are needed to 
understand
the true nature of CCOs.

\acknowledgements
This work was supported by NASA grant NAG5-10865 and SAO grants
AR3-4004B and GO3-4091X.  
We thank Slava Zavlin for useful discussions.

\end{document}